\documentstyle[aps,twocolumn,graphics,epsf,amsmath,amssymb,eqsecnum,floats]{revtex}
\begin{document}

\title{ The planar pyrochlore: a Valence Bond Crystal.}

\author{J.-B. Fouet, M. Mambrini, P. Sindzingre, C. Lhuillier}
\address{Laboratoire de Physique Th{\'e}orique des Liquides-UMR 7600
of CNRS,  Universit{\'e}  Pierre  et  Marie  Curie,  case 121,   4 place
Jussieu, 75252 Paris Cedex, France\\ E-mail:fouet@lptl.jussieu.fr}
\maketitle
\bibliographystyle{prsty}
(\today)\\
\begin{abstract}
Exact diagonalizations of the spin-1/2 Heisenberg model on the
checkerboard lattice have been performed for sizes up to $N=36$ in
the full Hilbert space and $N=40$ in the restricted subspace of first
neighbor dimers. This antiferromagnet does not break SU(2)
symmetry and displays long range order in 4-spin S=0
plaquettes. Both the symmetry properties of the spectrum and
various correlations functions are extensively studied. At
variance with the kagom{\'e} antiferromagnet, the Heisenberg quantum
model on a checkerboard lattice is a Valence Bond Crystal. Some
results concerning the 3-dimensional spin-1/2 pyrochlore magnet
(for sizes 16 and 32) are also shown: 
this system could behave differently from its 2-dimensional analog.
\end{abstract}

\section{INTRODUCTION}

In the family of frustrated magnets, the kagom{\'e} and pyrochlore
lattices have attracted special attention both experimentally
and theoretically. Experimentally such magnets display a
wide variety of unusual low temperature behaviors
\cite{r94,r00,rec90,ukkll94,kklllwutdg96,mklmch00}
signatures of different kind of 
collective low energy degrees of freedom.
 
The first-neighbor classical Heisenberg model on
such lattices has a T=0 entropy \cite{e89}.
On these lattices, the Heisenberg model can be rewritten as the sum of
the square of the total spin of corner sharing units $\alpha$
(triangles for the kagom{\'e} lattice, tetrahedra in the
pyrochlore):

\begin{equation}
{\cal H} =  J \sum_{(i,j)\,bonds} {\bf S}_i.{\bf S}_j
\equiv \frac {J}{2} \sum_{\alpha\,units} {\bf S_{\alpha}}^2 + Cst.
\label{eq-Heis}
\end{equation}
Thus a classical ground-state is obtained whenever ${\bf S_{\alpha}}=0$
for all $\alpha$.
It is a straightforward exercise to
show that such ground-states have a continuous local
degeneracy. Thermal fluctuations select planar spin
configurations on the kagom{\'e} lattice \cite{chs92,schb93},
but are unable to build order from disorder in the
pyrochlore lattice\cite{mc98}.

 A simple Maxwellian counting has been done by the last authors:
the number of degrees of freedom of $N$ Heisenberg spins with a given
length is $F=2N$.
The number of constraints  to realize a
classical ground-state is $K=6N/q$ where $q$
is the number of spins on each $\alpha$ unit. Assuming
that these constraints are linearly 
independent~\footnote{This has been argued to be true in the pyrochlore
case\cite{mc98}.}, one finds a $T=0$ extensive entropy ($F-K \sim N$)
for the pyrochlore  and zero entropy  ($F-K \sim 0$)
in the kagom{\'e} case. 
Although the assumption is known to fail for the kagom{\'e} magnet
this naive counting suggests  that the degeneracy of the classical 
ground-state in the pyrochlore magnet is larger than in kagom{\'e} magnet,
in qualitative agreement with
the thermal behavior of the two magnets. 

\begin{figure}
	\begin{center}
	\resizebox{6cm}{!}{\includegraphics{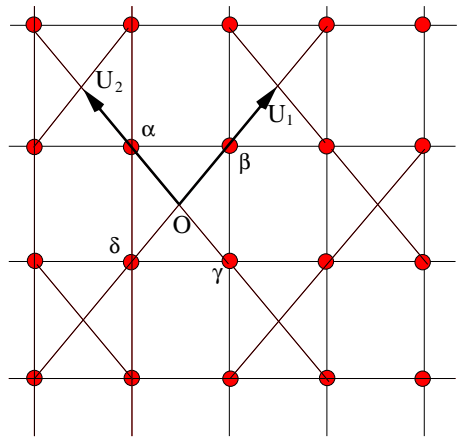}}  \end{center}

	\caption[99]{    The checkerboard lattice: the spins sit
at the vertices shown by bullets, all couplings are identical,
$\bf u_1, u_2$ are the unit vectors of the Bravais lattice.
	}  \label{checkerboard2}
\end{figure}

Lately, Palmer and Chalker have studied the
Heisenberg problem on the checker-board lattice\cite{pc00}.
 This lattice built out of corner sharing 4-spin squares
(see Fig.~\ref{checkerboard2}) is the two dimensional analog of the
pyrochlore lattice. The classical Heisenberg model 
on the checker-board lattice has a similar ground-state degeneracy 
and behave the same way at low temperature (and with additional 
dipolar interactions)\cite{pc00}.

The effect of quantum fluctuations on these different
structures remains to be fully understood. In the large S,
first-order spin-wave approximation, all these magnets 
remain disordered\cite{c01}. Higher order
approximations have been devised for the kagom{\'e} lattice
and lead to selection of order out of disorder by quantum
fluctuations\cite{c92}. There is no spin long range order (LRO) in the
pyrochlore magnet\cite{cl98,kk01}.

The spectrum of low lying excitations of the spin-1/2
kagom{\'e} magnet obtained from exact diagonalizations
has been a real surprise\cite{web98,smlbpwe00}: whereas it
probably has a small gap for $\Delta S =1$ excitations
(transitions $ S=0 \to S=1$), there is no gap to
singlet excitations (transitions $ S=0 \to S=0$)
and the density of low lying $S=0$ states is so large that
the system has a T=0 residual entropy. The discovery of a
second model with a similar spectrum of low lying
excitations on the triangular lattice with 4-spin exchange
interaction lead us to speculate that this could be a
generic new type of magnets\cite{lmsl00,lsf01}. A natural
question thus arises: do the 2-dimensional and the true
pyrochlore quantum magnets belong to this generic class? The
results obtained from their classical and semi-classical 
counterparts support the speculation that the answer might
be positive! As  exact diagonalizations are up to now
limited to systems of $N \sim 36$ spins, the problem of the
true spin-1/2 pyrochlore magnet might remain open for still
a long time. The 2-d pyrochlore looks more promising:
Palmer and Chalker\cite{pc01} 
have computed the spectra of
clusters up to $24 $ spins. From their results, they were
able to conclude that the system has no N{\'e}el LRO;
it does not break $SU(2)$ at $T=0$  and probably has
 a large spin gap.  Yet these sizes were not
large enough to be sure that this magnet was really in the
same class as the kagom{\'e} magnet. In this work we extend
such diagonalizations up to $N=36$.
The technical aspects of these diagonalizations
have been previously described\cite{bllp94}.

Besides these diagonalizations in the full Hilbert space,
we also have peformed diagonalizations in the restricted space of
first neighbor dimer coverings (denoted in the following FNSS, for
First Neighbor  Singlet Subspace). The size of this restricted
subspace is smaller than the $S=0$ sector of the full Hilbert
space and it increases slower with the system size ( $\sim 1.33^N$
compared to $\sim 2^N$). In this restricted basis we have studied
samples up to $N=40$.
The FNSS calculations for $N=40$ have required an order 
of magnitude less of computer memory than the full Hilbert space
calculations for $N=36$.
As usual, periodic boundary conditions are applied to the samples.

\begin{figure}
        \begin{center}
        \resizebox{8cm}{!}{
        \includegraphics{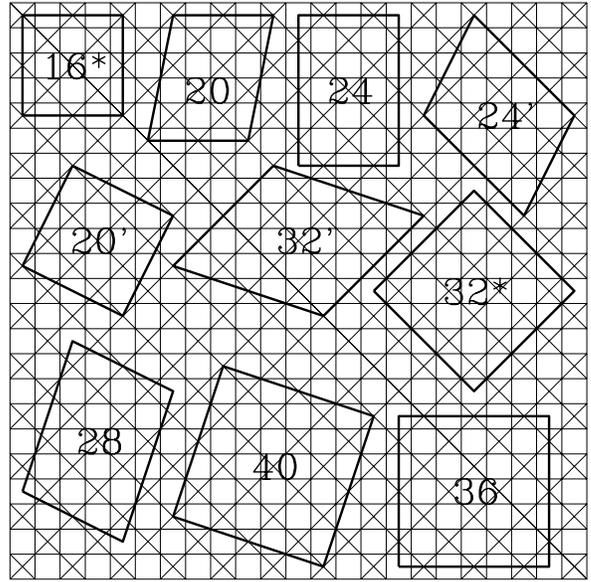}}
        \end{center}
\caption[99]{  Samples:
the star indicates that the sample has extra symmetries not shared
by the chekerboard infinite lattice (see text),
the prime indicates that the sample has higher ground-state energy
than the other with the same number of spins.
}
    \label{sqm_systems}
\end{figure}

\section { Energy per spin of the ground-state in the full
Hilbert space} 
              
\begin{figure}
        \begin{center}
        \resizebox{8cm}{!}{
        \includegraphics{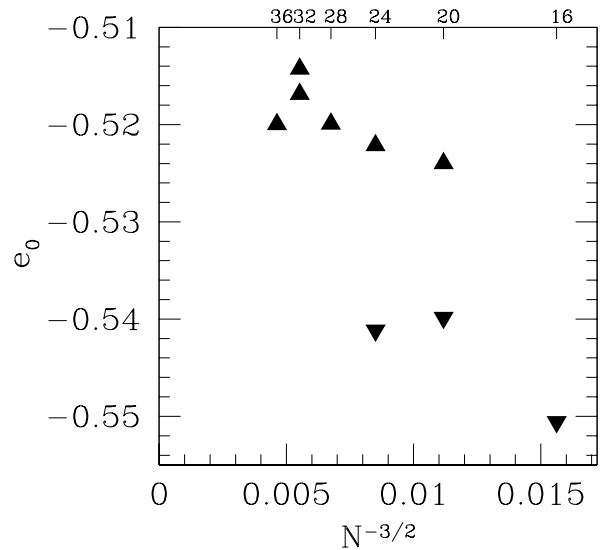}}
        \end{center}   
\caption[99]{  Energy per spin $e_0$ vs $N^{-3/2}$ for
``true 2-dimensional samples'' (full upwards triangles)
 and ``quasi 1-dimensional tubes''
 (downwards triangles)  (see text).
}
    \label{fig-e-per-spin-diago}
\end{figure}

\begin{table}
\begin{center}
\begin{tabular}{|c|c|c|c|c|c|c|c|c|}
\hline
  $N$ &  16*  & 20 & 24 &24'& 28 & 32*& 32' & 36 \\
  \hline
$e_0$&-.551&-.540&-.541&-.522&-.520&-.517&-.514&-.520\\
 \hline
  $E^1_{S=1}-E^1_{S=0}$ & 1.12 & 0.96 & 1.00 &0.58& 0.57 & 0.69 & 0.57& 0.71\\
  \hline
  $E^2_{S=0}-E^1_{S=0}$ & 0.37 & 0.30 & 0.20 &0.08& 0.09 & 0.03&0.01& 0.05 \\
  \hline
  $E^3_{S=0}-E^2_{S=0}$ &.53  & 0.22 & 0.28 &0.06& 0.05 & 0.18 &0.13& 0.22 \\
\hline
$E^1_{S=1}-E^3_{S=0}$&0.23&0.44&0.52&0.44&0.42&0.47&0.43&0.44\\
\hline
   $n_1$  & 27 &25 & 38 &51& 82 & 286&135& 110\\
\hline
$ln(n_1)$/N&0.21&0.16&0.15&0.16&0.16&0.18&0.15&0.13\\
\hline
\end{tabular}
\end{center} 
\caption[99]{ Spectrum of the Heisenberg model in the full Hilbert space.
Energy per spin in the ground-state $e_0$ and 
energy gaps $E^{n_S}_{S}-E^{n_S'}_{S'}$ between the ${n_S'}$ energy 
level
of the $S'$ spin sector and the ${n_S}$ level of the $S$ sector.
Second line: spin gap. Third line: gap between the absolute
 ground-state and the first singlet excitation . Fourth  line: gap
between the second and third level in the $S=0$ sector.
Fifth line: gap between the third level in the $S=0$ sector
 and the first triplet excitation.
Following lines: 
 $n_1$ is the number of singlet
states in the spin gap (including degeneracies). 
The starred columns correspond to samples which have
the extra symmetries of the pyrochlore lattice.
The three first columns are 4-spins tubes.
}
\label{table1}
\end{table}

The absolute ground-state is an S=0 state\cite{liebschupp99},
in the trivial representation of the space group.
The ground-state energy per spin versus system size 
is given in Fig.~\ref{fig-e-per-spin-diago} and in Table~\ref{table1}.
The samples are displayed in Fig.~\ref{sqm_systems}.
Our results are identical to those of Palmer and Chalker\cite{pc01}
for the small sizes and identical shapes. 
 We have added some extra shapes (indexed by a prime) to show
the sensitivity of  small size results to the shape. 
The analysis of the whole set of results shows that
the  most stable small samples largely overestimate the
thermodynamic binding energy. In fact the first three samples
16, 20, 24 can be seen as tubes with a 4-spin section. The
properties of these quasi 1-dimensional systems is different from those of
the true 2-dimensional samples (see section VI). This is manifest in Fig.3 and
following and has also been checked on the properties of the
ground-state wave-function.

The energy per spin for the largest sizes seems to
 level off in the range $[-0.52,-0.51]$ .

The still non negligible size effect found on
samples 28, 32 and 36 has to be related to symmetry problems:
the $28$ sample has not all the symmetries of
the infinite lattice and the $32^{\ast}$ sample 
(as the $16^{\ast}$ sites sample)
 has extra
symmetries not shared by the  checkerboard infinite lattice (see below).
So in all respects
the $36$ sample seems the better sample to mimic the checkerboard
infinite lattice: its energy gives a plausible lower bound of the
thermodynamic limit\footnote{Its only weakness could be the absence of
fluctuations at wave-vectors $(0,\pi),(\pi,0)$, but all the 
information gathered on this system lead us to conclude that
this absence is not qualitatively essential.}.

Samples $16^{\ast}$ and $32^{\ast}$ are peculiar. 
There is in fact a one to one mapping, preserving  neighborhood
relationships and periodic boundary conditions,
between these samples  on the checkerboard lattice
and the cells of the
same size of a pyrochlore. Their symmetry group has extra
symmetries inherited from those of the pyrochlore lattice.
This explains the
extra  degeneracy noticed on exact spectra of these "pyrochlore" samples,
 when analyzed with the  checkerboard symmetry group~\footnote{Numerical
results available on request at fouet@lptl.jussieu.fr or phsi@lptl.jussieu.fr}. 

\section {Ground-state in the first-neighbor dimer subspace}

\begin{figure}
	\begin{center}
	\resizebox{8cm}{!}{\includegraphics{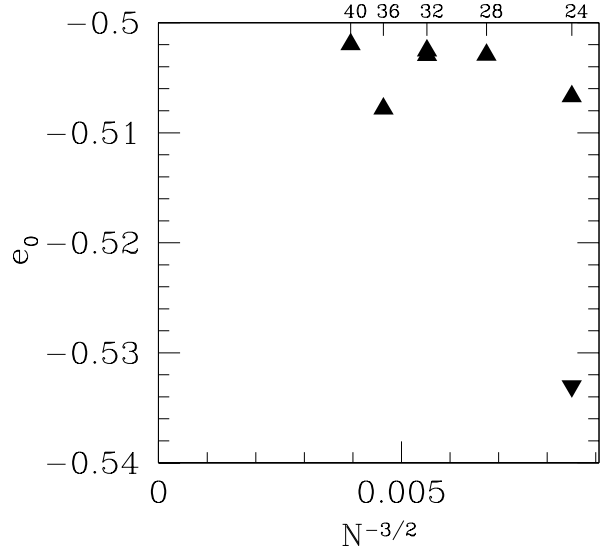}} 
 \end{center}

	\caption[99]{ Energy per spin in the FNSS subspace.
Symbols are the same as in Fig.~\ref{fig-e-per-spin-diago}. The
two $N=32$ results are nearly indistinguishable with the scale of the
symbols.
	}  \label{energy_FNSS}
\end{figure}

\begin{table}
\begin{center}
\begin{tabular}{ |c|c|c|c|c|c|c|c|}  \hline
   N     &24&24'&28&32*&32'&36&40\\  \hline
$E^2_{S=0}-E^1_{S=0}$&0.340&0.080&0.035&0.025&0.027&0.050&0.014\\ \hline
$E^3_{S=0}-E^2_{S=0}$&0.47&0.35&0.31&0.45&0.44&0.49&0.45\\ \hline
$(E_{var}-E_{ex})/E_{ex}$&0.015&0.029&0.008&0.027&0.023&0.023&\\ \hline
\end{tabular}
\end{center}
\caption[99]{Spectrum of the Heisenberg Hamiltonian in the first
neighbor singlet subspace. First two lines: energy gaps in the
singlet sector (same definitions as in Table 1). Last line:
relative difference in ground-state energy between the FNSS and
the full Hilbert space.}
\label{spect_FNSS}
\end{table}

Results of diagonalizations in the first neighbor singlet subspace
(Fig.~\ref{energy_FNSS} and Table \ref{spect_FNSS})
confirm  the above-mentioned hypotheses and call for the
following comments:
\begin{itemize}
\item The variational energy in the FNSS is $\sim 2\%$ above the
exact one. This property is not spoiled by increasing  system size. We
might thus expect that this variational subspace capture most of
the physics of the exact ground-state.
\item The size and shape effects on the $S=0$ ground-state are
roughly the same in the two sets of results. Due to the partial
cut-off of long dimers the  size effects in the FNSS are smaller
than in the exact ground-state.
\item The anomaly of the N=32 sample is less pronounced in the
FNSS.  This can be understood as this basis does not allow
expression of the full ternary symmetry of the 3d pyrochlore.
In fact the difference in ground-state energy between the two
different 32 samples is  hardly visible on the scale of
Fig.~\ref{energy_FNSS}.
\item The 40 sites sample 
has an energy in the same range as the
N=28,32,36 samples confirming that the larger sizes are in a
cross-over regime, with linear dimensions of the order of, or
larger than  the spin-spin correlation length.
\end{itemize}

\section{Spin gap}

\begin{figure}
	\begin{center}
	\resizebox{8cm}{!}{\includegraphics{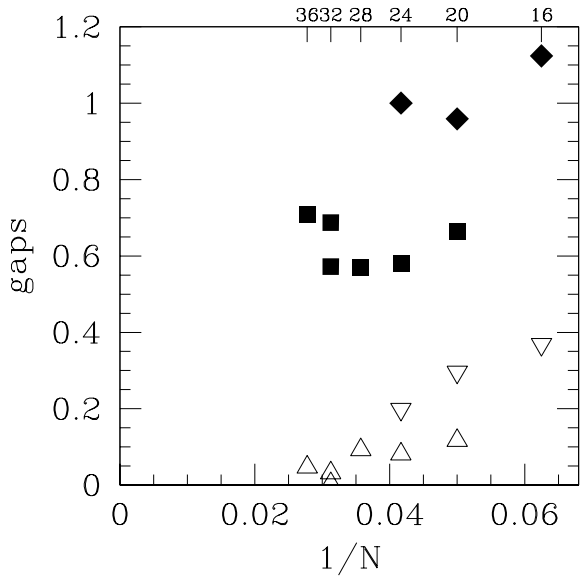}}
	\end{center}
\caption[99]{ Full Hilbert space: energy gaps measured from the
 absolute ground-state versus $1/N$. Full squares (diamonds):
spin-gaps for  2-dimensional samples (4-spin tubes). Open triangles
pointing up (down): gaps to the $2^{nd}$ singlet energy level
for 2-dimensional samples (4-spin tubes).}
    \label{fig-gaps-diago}
\end{figure} 
 
\begin{figure}
	\begin{center}
	\resizebox{8cm}{!}{\includegraphics{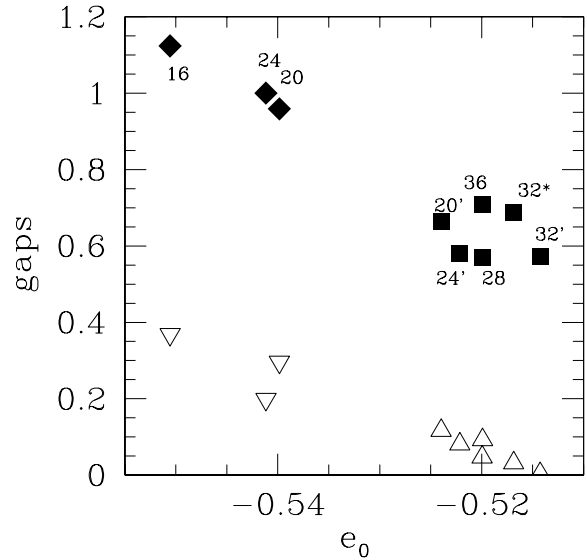}} 
	 \end{center}
	\caption[99]{Full Hilbert space: correlations between gaps and 
ground-state energy per site. Same symbols as in Fig.~\ref{fig-gaps-diago}.}
	  \label{gge3}
\end{figure}

The spin gap 
 (defined as the difference in total energy between 
the first $S=1$ excited state and the $S=0$ absolute ground-state)
is displayed in Fig.~\ref{fig-gaps-diago} and
Table~\ref{table1} (line 2) versus system size, and in Fig.~\ref{gge3}
versus ground-state energy per spin. This last figure emphasizes the
difference between the quasi 1-dimensional systems (tubes) with a
large binding energy ($\sim -0.54$) and a large gap ($\sim 1$),
and the true 2d systems with a binding energy ($\sim -0.52$) and a
gap of the order of 0.6 times the coupling constant.

\section{ Spectrum of the first excitations in the $S=0$ sector}

\begin{figure}
	\begin{center}
	\resizebox{8cm}{!}{\includegraphics{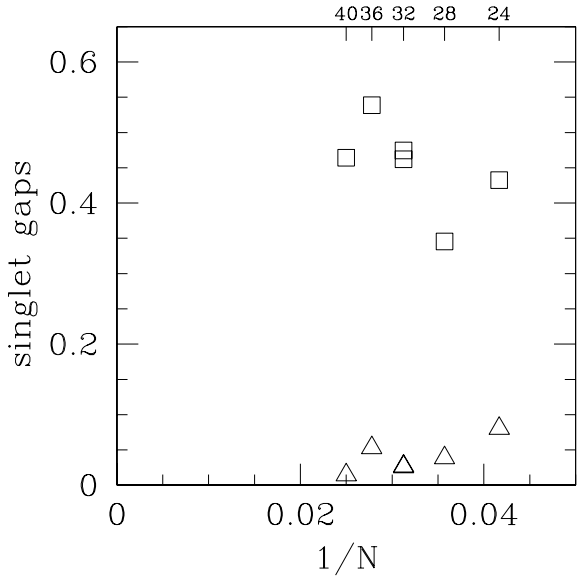}} 
 \end{center}

	\caption[99]{Gaps in the singlet sector. Open
triangles (open squares): gap from the ground-state to the first
(second) excited state in the FNSS.
	}  \label{gaps_FNSS}
\end{figure}

The first excited state in the full singlet sector collapses to the
ground-state with increasing system size (third line of Table 1,
open triangles in Figs.~\ref{fig-gaps-diago} and \ref{gge3}).
The same phenomenon is clearly seen in the FNSS (Table 2 and
Fig.~\ref{gaps_FNSS}). This is a clear indication of a degeneracy
of the absolute ground-state in the thermodynamic limit. In this
system with two spins 1/2 per unit cell we do not expect a
topological degeneracy\cite{ml00}.
We will explicit in the next section the space  symmetry breaking 
at the origin of the present degeneracy.

Analysis of the gap between the second and third singlet (fourth
line of Table 1 in the full Hilbert space, 2nd line of Table 2
in the FNSS and Fig.~\ref{gaps_FNSS}) shows a non monotonous
behavior and no tendency to close for larger sizes.
On the basis of the present results one expects, in the thermodynamic limit,
a finite gap in the singlet sector above the
2-fold degenerate ground-state  and this gap is 
probably smaller than the gap to the first
triplet (fifth line of Table 1 to be compared to the second
line of this same table).

Also shown in Table 1  is the number $n_1$ of singlet states in the spin-gap.
The unusually large values of $n_1$ found for the small samples
were taken by Palmer and Chalker\cite{pc01}
as indications of a similarity of the checkerboard
and the kagom{\'e} magnet.  These large values are probably an indication 
of a continuum of singlet excitations.
But this continuum appear to be separated from the ground-sate by a finite gap.
So the continuous density of singlet states adjacent to the
ground-state, that is the distinctive characteristic of the
kagom{\'e} magnet, is absent in  the
2-dimensional checkerboard lattice (Fig.~\ref{gaps_FNSS}).

Endly the continuum of singlet excitations above the third level
of the true checker-board spectra can be interpreted as the
excitations of antiferromagnetic pairs of confined spinons. The
ferromagnetic pairs appear more energetic (fifth line of Table 1
to be compared to the second line).

 In view of the results
for the $N=16^{\ast}$ and $32^{\ast}$ ``pyrochlore'' samples
one might speculate a different behavior for the pyrochlore lattice:
$n_1$ is indeed much larger than for other sizes (Table.~\ref{table1}).
But no continuum is yet actually visible in the spectrum as it is in the 
kagom{\'e} spectrum.
 There is still a noticeable gap between the second and third singlet
eigenlevels in the $32^{\ast}$ sample. 
Larger sizes would be necessary to really see if the
pyrochlore belongs to the same generic class as the kagom{\'e}.

\section{ Ground-state symmetry breaking}
\begin{figure}
	\begin{center}
	\resizebox{8cm}{!}{\includegraphics{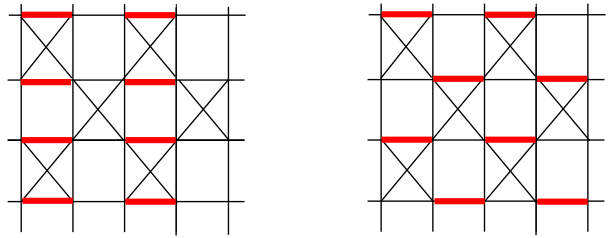}}  \end{center}

	\caption[99]{ Columnar and staggered configuration of
dimers (fat links) on the checkerboard lattice: such symmetry breaking
 configurations are 4-fold degenerate in the thermodynamic limit.}
	  \label{dimVBC}
\end{figure}

As noticed above the finite size results point to a
2-fold degeneracy of the ground-state in the
thermodynamic limit. The absolute ground-state is in the
trivial representation of the lattice symmetry group. Its
wave function is invariant in any translation and
 in any operation of $D_4$: group
of the $\pi /2$ rotations around  point O (or any equivalent point
 of the Bravais lattice)  and axial symmetries with respect to axes 
$\bf u_1$ and $\bf u_2$ (see Fig.~1).
  The excited
state which collapses on it in the thermodynamic limit
 has a wave vector $(\pi,\pi)$ (its wave function takes a
(-1) factor in one-step translations along  $\bf u_1$ or $\bf u_2$), and it is
odd under $\pi /2$ rotations 
and axial symmetries.
In the thermodynamic limit the 2-fold degenerate ground-state
can thus exhibit a spontaneous symmetry breaking with a doubling
of the unit cell. Such a restricted  symmetry breaking does
not allow a columnar or staggered configuration of dimers  
(Fig.~\ref{dimVBC}): both of these states have at least a 4-fold
degeneracy. 

\begin{figure}
	\begin{center}
	\resizebox{8cm}{!}{\includegraphics{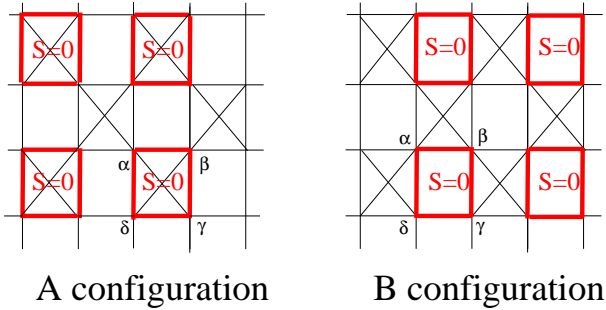}}  \end{center}

	\caption[99]{    S=0 4-spin plaquette  valence-bond   crystals
  on the  checkerboard lattice:  fat links indicate  4 spins
involved in                 a            singlet.
	}  \label{plaqVBC}
\end{figure}

 The simplest Valence Bond Crystals  that allow the above-mentioned 
symmetry breaking  are described by pure product wave-functions 
of 4-spin  S=0 plaquettes. This family includes eight different
configurations: 
\begin{itemize}
\item
The singlet plaquettes
  may sit either on the squares with crossed links or on
the void squares (A and B configurations of Fig.~\ref{plaqVBC}),
\item  The translation symmetry breaking configurations may be in
two different locations named $A_{1(2)}$ (resp $B_{1(2)}$),
\item An S=0 state on a plaquette of four spins sitting on sites
($\alpha, \beta, \gamma, \delta$) may be realized either by
the symmetric combination of pairs of singlets: 
\begin{equation}
|\psi^{+}> = |\alpha\to\delta>|\gamma\to\beta> + 
|\alpha\to\beta>|\gamma\to\delta>,
\label{splaq}
\end{equation}
 or by the anti-symmetric one:
\begin{equation}
|\psi^{-}> = |\alpha\to\delta>|\gamma\to\beta> -
|\alpha\to\beta>|\gamma\to\delta>.
\label{aplaq}
\end{equation}
where $|\alpha\to\gamma>$ is the singlet state on sites 
$\alpha$ and $\gamma$:
\begin{equation}
|\alpha\to\gamma>  = 
(|\alpha \uparrow, \gamma\downarrow> -
|\alpha\downarrow, \gamma\uparrow>)/\sqrt2. 
\end{equation} 
\end{itemize}

\begin{table}
\begin{center}
\begin{tabular}{|c||c|c|c|}
Wave-function& ${\cal T}_{{\bf u}_1}$& ${\cal R}_{\pi /2}$&$ {\cal
\sigma}_{{\bf u}_1}$\\
\hline
\hline
$A_{1(2)}^{+}$&$A_{2(1)}^{+}$&$A_{1(2)}^{+}$&$A_{1(2)}^{+}$\\
$A_{1(2)}^{-}$&$A_{2(1)}^{-}$&$(-1)^pA_{1(2)}^{-}$&$(-1)^pA_{1(2)}^{-}$\\
$B_{1(2)}^{+}$&$B_{2(1)}^{+}$&$B_{2(1)}^{+}$&$B_{2(1)}^{+}$\\

$B_{1(2)}^{-}$&$B_{2(1)}^{-}$&$(-1)^pB_{2(1)}^{-}$&$(-1)^pB_{2(1)}^{-}$\\
\hline
$X^{\eta}= A_{1}^{+} +\, \eta \, A_{2}^{+}$&$ \eta\, X^{\eta}$&$ X^{\eta}$
&$X^{\eta}$\\
$Y^{\eta}= A_{1}^{-} +\, \eta \,A_{2}^{-}$&$ \eta\, Y^{\eta}$&
$(-1)^p Y^{\eta}$&$(-1)^p Y^{\eta}$\\
$Z^{\eta}= B_{1}^{+} +\, \eta \, B_{2}^{+}$&$ \eta\, Z^{\eta}$&
$\eta\, Z^{\eta}$&$\eta\, Z^{\eta}$\\
$T^{\eta}= B_{1}^{-} +\, \eta \, B_{2}^{-}$&$ \eta\, T^{\eta}$&
$(-)^p \,\eta\, T^{\eta}$&$(-)^p\, \eta\, T^{\eta}$\\
\end{tabular}
\end{center} 
\caption[99]{Transformation rules of the product wave-functions
in the elementary operations of the symmetry group
(the space group is defined  with respect to point O
and translations ${\bf u_1, u_2}$).
The wave-functions of the anti-symmetric plaquettes 
have different symmetries depending on the parity p of the number
of plaquettes in the sample.}
\label{symmetryop}
\end{table}

We can thus define eight different product wave-functions
labeled: $|A_{1(2)}^{\epsilon}>$ and $|B_{1(2)}^{\epsilon}>$.
The transformations of these states under the elementary
operations of the  lattice symmetry group are described in
the first four lines of 
Table \ref{symmetryop}.
The symmetric (resp. anti-symmetric) linear combinations of these states
which are irreducible representations of this group are defined
in the four last lines of the same Table.
The comparison of the symmetries of these states with those of
the two first levels of the exact spectra indicates a Z type 
symmetry  of the checker-board magnet ground-state: in the 
thermodynamic limit the symmetry
breaking configuration is thus of the B type decorated by the
symmetric 4-spin plaquettes described in Eq.\ref{splaq}
 (anti-symmetric 4-spin plaquettes  are excluded by the
 properties of the exact ground-state and first singlet excitation
 in samples with an odd number
of 4-spin plaquettes, such as  N=28 or 36).

Indeed in the exact ground-state quantum fluctuations 
 might  dress the product states giving a more fuzzy picture.
Insofar as gaps to the third singlet state and the first triplet
remain finite in the thermodynamic limit, the Valence Bond Crystal picture
 (with LRO in plaquettes) will survive to quantum
fluctuations.

A simple last remark could be done: the
symmetric-plaquette state (Eq.~\ref{splaq})
can be rewritten as the product of two
triplets along the diagonals of the square. This
configuration of spins is not energetically optimal on the squares with
antiferromagnetic crossed links  (A configuration)
but might a priori be favored in
B configuration. Reversely the $\psi^{-}$-plaquette can be rewritten
as the product of two
singlets along the diagonals of the square, and would
eventually be preferred in A configuration.
The variational energy per spin of the product wave-function of
$\psi^{+}$
plaquettes in B configuration  is $E_{var}(B^+)= -0.5$, whereas the
 variational energy per spin of the product wave-function of $\psi^{-}$
plaquettes in A configuration is $E_{var}(A^-)= -0.375$.
For the quasi 1-dimensional samples 16,20 and 24, 
these symmetric plaquettes can be built not only on square voids but
also along the 4-spin cross section of the tube.
The resonnance between plaquettes on void square and plaquettes on the
cross section   might explain the special properties (energy,
gaps, correlations) of these samples.  

 Exact results are indeed consistent with this variational
estimate and favor $\psi^{+}$-plaquettes on voids. This is in 
agreement with recent results of Moessner $et \; al$~\cite{mts01}
but is at variance with the departure point of the strong coupling
approximation of Elhajal $et \; al$~\cite{ecl01}.

\section{ Correlations}

\begin{table}
\begin{center}
\begin{tabular}{|c|c|c|c||c|c|c|}
${i,j}$& ex. g.-s.&Z w-f.& FNSS&${i,j}$& ex. g.-s.&Z w-f.\\
\hline
{1,2} &-0.239 &-0.25 &-0.27  &{1,29}& 0.001 &0.  \\
{1,8} &-0.043 & 0.   & 0.033 &{1,17}&-0.002 &0.  \\
{1,32}& 0.088 & 0.125& 0.122 &{1,4 }&-0.037 &0.  \\
{1,3} & 0.034 & 0.   &       &{1,10}&-0.012 &0.  \\
{1,35}& 0.013 & 0.   &-0.018 &{1,16}&-0.001 &0.  \\
\end{tabular}
\end{center}
\caption[99]{ Spin-Spin correlations
${\cal C}^2 (i,j)=<\bf{S}_i.\bf{S}_j>$ in the exact ground-state
(second columns) , in the variational Z
 wave-function (third columns), and in the ground-state of the
 first neighbor singlet subspace (fourth columns) of the N=36 sample.
 The sites $i,j$ are numbered as in Fig.~\ref{corr}.}
\label{spin-spincorr}
\end{table}

As it was expected for a Valence Bond Crystal it can be seen in
  Table~\ref{spin-spincorr} that
  spin-spin correlations in the $N=36$ exact 
ground-state decrease very rapidly with distance. The decrease
with distance is
even more rapid in the full Hilbert space than in the FNSS. In
this respect the Z product wave-function appears to be a good
simple variational guess to describe the exact ground-state.

\begin{figure}
	\begin{center}
	\resizebox{8cm}{!}{\includegraphics{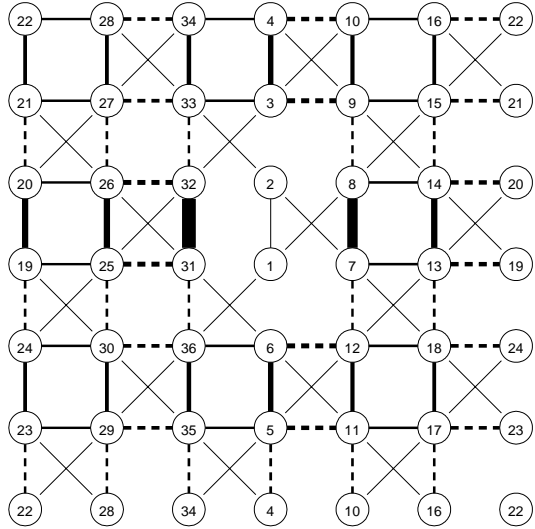}} 
 \end{center}

	\caption[99]{ Dimer-dimer correlations in the exact
ground-state of the 36 sample (Eq.~\ref{dim-dim-corr}).
The reference bond  is the bond $(1,2)$. Positive
(negative)
correlations are drawn as full (dashed) lines. The thickness of the
lines is a measure of the strength of the correlation.
The diagonal lines show the position of the crossed links.
	}  \label{corr}
\end{figure}

The 4-point correlation function:
 \begin{equation}
{\cal C}^{4}(1,2;i,j)= 4 \left[< {\bf S}_1.{\bf S}_2\;{\bf S}_i.{\bf S}_j>
 - < {\bf S}_1.{\bf S}_2><{\bf S}_i.{\bf S}_j>\right]
\label{dim-dim-corr}
\end{equation}
 is displayed in Table~\ref{dimer-dimercorr} in the exact
ground-state, in the Z product wave-function and in the
ground-state of the First Neighbor Singlet Subspace (see also
Fig.~\ref{corr}). 
Here again the general behaviors are
 quite similar. As it was expected quantum fluctuations in the 
exact g.-s. renormalize the correlations at intermediate and larger
 distances. Asymptotic behavior seems approximately reached for
 the larger distances in sample $N=36$. 
Renormalization  by quantum fluctuations amounts to  $\sim 60\%$
 of the bare variational Z value. 

Due to the plaquette structure of this Valence Bond Crystal a better
order parameter is given by the cyclic permutation operator
$P_{\alpha, \beta, \gamma, \delta}$ of the
4 spins $(\alpha, \beta, \gamma, \delta)$ on the square plaquette.
Let us define the corresponding hermitic observable as:
\begin{eqnarray}
Q_{\alpha, \beta, \gamma, \delta}& =& 
\frac {1}{2} ( P_{\alpha, \beta, \gamma, \delta} 
            +P^{-1}_{\alpha, \beta, \gamma, \delta})\nonumber\\
 &=& 2 \left[ 
{\bf S}_{\alpha}.{\bf S}_{\beta}\;{\bf S}_{\gamma}.{\bf S}_{\delta}
+{\bf S}_{\alpha}.{\bf S}_{\delta}\;{\bf S}_{\beta}.{\bf S}_{\gamma}
-{\bf S}_{\alpha}.{\bf S}_{\gamma}
\;{\bf S}_{\beta}.{\bf S}_{\delta} \right]\nonumber\\
&& +0.5 \left[  {\bf S}_{\alpha}.{\bf S}_{\beta} + 
{\bf S}_{\gamma}.{\bf S}_{\delta} + 
{\bf S}_{\alpha}.{\bf S}_{\delta} + 
{\bf S}_{\beta}.{\bf S}_{\gamma}\right]\nonumber\\
&&+0.5\left[ {\bf S}_{\alpha}.{\bf S}_{\gamma} +
{\bf S}_{\beta}.{\bf S}_{\delta} +1/4\right]
\label{4-spin-obs}
\end{eqnarray} 

Its value in the N=36 exact ground-state (resp. in the Z variational
 w.-f.) is 0.478 (resp. 0.56) on a void square, and 0.071 (resp. 0.125) 
on a square with crossed links. The correlation function
of this observable is defined as usual as: 
\begin{eqnarray}
{\cal C}^8 (\alpha, \beta, \gamma, \delta ; i,j,k,l)&=& 
<Q_{\alpha, \beta, \gamma, \delta} Q_{i,j,k,l}>\nonumber \\
&&- <Q_{\alpha, \beta, \gamma, \delta}><Q_{i,j,k,l}>
\label{4-spin-corr}
\end{eqnarray}
Its values are displayed in Table~\ref{plaq_corr}.
One might notice the presence of non negligible correlations
 between void squares and the quasi absence between squares with
 crossed links. The short distance value of these correlations
 shows the limits of relevance of the variational description.
As expected from the spectra, the checkerboard Heisenberg magnet is
 a Valence Bond Crystal.

\section{Conclusion}

The spin-1/2 checkerboard Heisenberg antiferro-magnet is a Valence Bond
Crystal with LRO in 4-spin S=0 plaquettes: it exhibits a large
spin gap, a breaking of the translational symmetry, a doubling
of the unit cell and long range correlations in singlets. 

At variance with the kagom{\'e} antiferromagnet this system does not
 exhibit a singlet continuum but a clear gap in the singlet sector
above the (quasi-)degenerate ground-state.

This system is a 2-dimensional analog of the dimerized phase of the
$J_1-J_2$ on a chain. It belongs to the same generic class as 
the Shastry-Sutherland model\cite{cms01} 
and the $J_1-J_2$ model\cite{cs00,sow00,jongh_leeuwen00} on the
square lattice for $J_1-J_2 \sim 0.5$.
Dimer LRO has also been found
on the $J_1-J_2$ model on the honeycomb lattice\cite{fsl01}. 
In these last examples,
the dimerized phases are found after
destabilization of a classical collinear N{\'e}el ground-state
by quantum fluctuations as predicted 
from SU(N) or Sp(N) approaches~\cite{rs90,read-sachdev91}.
A common feature of all these magnets is a bipartite lattice. 
Such an underlying lattice is probably favorable for the
establishment of LRO in dimer coverings.
\footnote{ In each of these models the
system is around the point of maximum classical frustration
obtained for $z_2 J_2 = z_1 J_1/2$ (with $z_i$ and $J_i$
are respectively the coordinance and the coupling  at distance $i$).
In the $J_1-J_2$ model on the square lattice there
are still some controversies on the exact nature of the singlets
with LRO (dimers or 4-spin
plaquettes)\cite{cs00,sow00,jongh_leeuwen00},  but no doubt
about the belonging of this phase to the Valence Bond Crystal
family.}

As shown in this paper the physics of the model in the singlet
sector can essentially be captured in the restricted space of
first neighbor coverings (FNSS). This explains why the Quantum
Hard Core Dimer model on such a lattice gives essentially the same
physics and phase diagram as the full Heisenberg model\cite{mts01}.
Nevertheless the renormalization by quantum fluctuations in the
full Hilbert space is somewhat underestimated in the FNSS, and 
{\it a fortiori} in the Quantum Hard Core Dimer model.

This work also brings a new light on the discussion about the
kagom{\'e} magnet. The local classical degeneracy of a model
(present in any system with corner sharing units -as discussed in
the introduction) is not  a sufficient condition for the
associated quantum model to exhibit a continuum of singlets and a
residual entropy. Two pieces of information join to cast a doubt
on the relationship between a classical continuous degeneracy and
a kagom{\'e} like spectrum 
(type II Resonating Valence Bond Spin Liquid \cite{lm01}):
\begin{itemize}
\item The checkerboard magnet has a continuum degeneracy in the
classical limit but no continuum of singlets.
\item The multi-spin exchange model on the triangular lattice
(with an antiferromagnetic first neighbor coupling) has a
continuum of singlets in the triplet gap but apparently no simple
local continuous degeneracy in the classical limit.
\end{itemize}

Endly our results seem to indicate qualitative differences between
planar and true 3-d pyrochlore (see section V above).
In fact in the 3-d pyrochlore magnet,
symmetric and antisymmetric spin
singlets configurations on the tetrahedra are degenerate (as in the
checker-board lattice), but there are no unique 4-spin S=0
configurations around the ``voids'' of the structure and the
number of resonances on the loops encircling these voids is
large.
These quantum resonances are probably a very efficient mechanism
to destabilize dimer LRO. 

\begin{table}
\begin{center}
\begin{tabular}{|c|c|c|c||c|c|c|c|}
${i,j}$& ex. g.-s.&Z w-f.& FNSS&${i,j}$& ex. g.-s.&Z w-f.& FNSS\\
\hline
{31,32}&.56&.63&.55&{7,13}&.10&.25&.16\\
{7,8}&.43&.42&.39&{19,25}&.10&.25&.14\\
{25,26}&.26&.25&.21&{7,12}&-.10&-.25&-.15\\
{13,14}&.26&.25&.21&{31,36}&-.10&-.25&-.15\\
{19,20}&.25&.25&.21&{13,18}&-.11&-.25&-.16\\
{6,5}&.22&.25&.26&{25,30}&-.11&-.25&-.15\\
{6,12}&-.20&-.25&-.18&{19,24}&-.11&-.25&-.15\\
{25,31}&-.20&-.25&-.18&{6,36}&.10&.25&.16\\
{13,19}&-.18&-.25&-.18&{12,18}&.11&.25&.16\\
{36,35}&.18&.25&.18&{24,30}&.10&.25&.15\\
{5,11}&-.18&-.25&-.18&{35,5}&.10&.25&.15\\
{4,10}&-.18&-.25&-.18&{11,17}&.10&.25&.15\\
{12,11}&.17&.25&.19&{29,23}&.10&.25&.14\\
{36,30}&-.15&-.25&-.16&{5,4}&-.11&-.25&-.16\\
{35,29}&-.15&-.25&-.16&{11,10}&-.11&-.25&-.14\\
{30,29}&.15&.25&.17&{35,34}&-.11&-.25&-.14\\
{17,23}&-.15&-.25&-.16&{17,16}&-.11&-.25&-.14\\
{18,17}&.15&.25&.17&{29,28}&-.10&-.25&-.14\\
{18,24}&-.15&-.25&-.16&{23,22}&-.10&-.25&-.14\\
{24,23}&.15&.25&.16&{34,4}&.10&.25&.15\\
{28,34}&-.15&-.25&-.16&{10,16}&.10&.25&.15\\
{16,22}&-.15&-.25&-.16&{28,22}&.10&.25&.14\\
\end{tabular}
\end{center}
\caption[99]{ Dimer-dimer correlations ${\cal C}^4(1,2;i,j)$ (Eq.
\ref{dim-dim-corr}) in the $N=36$ ground-state. 
 The sites $1,2,i,j$ are described in Fig.~\ref{corr},
the $i,j$ points are enumerated in the first columns.
 This correlation has been measured in the exact ground-state
 wave function ( second columns), in the variational Z state 
(third columns) and in the ground-state of Eq.~1 
 in the first neighbor singlet subspace (FNSS, fourth columns).
All the values of these correlations between sites of Fig.~\ref{corr}
 can be obtained from this table by a mirror symmetry through the 
bisector of bond $(1,2)$.}
\label{dimer-dimercorr}
\end{table}

\begin{table}
\begin{center}
\begin{tabular}{|c|c||c|c|}
(1,31,32,2;  5,35,36, 6)& .172&(7,1,2,8; 11, 5 ,6,12)& .073\\
(1,31,32,2; 17,11,12,18)& .127&(7,1,2,8; 23,17,18,24)& .009\\ 
(1,31,32,2; 10, 4, 5,11)&-.121&(7,1,2,8; 16,10,11,17)&-.006\\
(1,31,32,2; 22,16,17,23)&-.117&(7,1,2,8; 28,22,23,29)&-.004\\
\end{tabular}
\end{center}
\caption[99]{Plaquette-plaquette correlations
${\cal C}^8 (\alpha, \beta, \gamma, \delta ; i,j,k,l)$ (Eq.~\ref{4-spin-corr})
in the exact ground-state of the $N=36$ sample. The sites are numbered
as in Fig.~\ref{corr}. The left part of the table describes the
correlations between void squares, the right part between squares with
crossed links.}
\label{plaq_corr}
\end{table}

\end{document}